\documentclass[12pt]{article}

\usepackage{graphicx}
\usepackage{latexsym,amsfonts, amssymb,epsfig}
\usepackage[english]{babel}

\begin{document}
\begin{center}
{\Large {\bf On the stability of some exact solutions to the
generalized convection-reaction-diffusion equation}

\vspace{2mm}

{\it V. Vladimirov and Cz. M\c{a}czka} } \vspace{2mm}

{Faculty of Applied Mathematics, \\
  AGH University of Science and Technology, \\
Mickiewicz Avenue 30, 30-059 Krak\'{o}w, PL} \\
Email address: {vsevolod.vladimirov@gmai.com}
\end{center}

\vspace{2mm}

 {\footnotesize {\bf

Abstract}
 Stability of a set of travelling wave solutions
to the hyperbolic generalization of the
convection-reaction-diffusion equation is studied by means of the
qualitative methods and numerical simulation. }



\section{Introduction}

In recent decades significant attention  was paid to the study of
the family of convection-reaction-diffusion equations
\begin{equation}\label{CRD}
u_t=\left[\kappa(u)\,u_x \right]_{x}+a(u)\,u_{x}+f(u).
\end{equation}
Equations belonging to this family describe a number of natural
phenomena, such as transport in porous media, or the motion of a
thin sheet of viscous liquid over the inclined plate (see
\cite{Kersner} and the literature therein). This class also contains
a nonlinear generalization of the Focker-Plank equation
\cite{Richards} and a number of models encountered in the biological
sciences \cite{Murray}. Another source of inspiration for studying
the convection-reaction-diffusion equations results from the fact
that the equation (\ref{CRD}) represents one of the simplest
nonlinear models describing phenomena of patterns formation and
evolution \cite{Prigogine,Haken}. It is, perhaps, the combination of
relative simplicity and richness of physical contents, that made the
family (\ref{CRD}) the objective of numerous studies within the
symmetry approach, purposed at constructing  nontrivial exact
solutions and finding out the conservation laws
\cite{Bar_Yur}--\cite{Vlask_04}.

In this paper, we consider the following evolutionary
 equation (referred to as GBE):
\begin{equation}\label{GBE1}
\alpha\,u_{tt}+u_t+\mu\,u\,u_x-\kappa\,u_{xx}=f(u).
\end{equation}
Here $\mu,\,\kappa$ are positive constants, $\alpha$ is nonnegative,
$f(u)$ is a smooth (polynomial) function, which will be specified
later on. Equation (\ref{GBE1}) is a hyperbolic generalization of
 the convection-reaction-diffusion equation. Let us
note, that the term $\alpha\,u_{tt}$ appears when the memory effects
are taken into account \cite{Joseph}--\cite{Kar}. Equation
(\ref{GBE1}), as well as its numerous modifications, were intensely
studied in recent years within the generalized symmetry approach
\cite{Vlask_04}, \cite{Kar}--\cite{Vlamacz_10}. Owing to these
studies, the analytical description of a large variety of traveling
wave (TW) solutions is actually available, including interacting
traveling fronts, soliton-like solutions, periodic waves,
compactons, shock fronts and many other. Undoubtedly, knowledge of
exact solutions to a non-linear PDE is a great advantage. At the
same time, individual exact solution is interesting and important
from the point of view of applications, if it is typical in some
sense to the equation under consideration. In most noteworthy cases,
self-similar exact solutions serve as the intermediate (or the true)
asymptotics \cite{Barenblatt}--\cite{Kamin_Rosenau2}, manifesting
attracting features.

The first stage towards the estimation of validity of the exact
solution is a study of its stability, and this is the main topic of
the present work. We formulate the conditions  which guarantee the
stability of some class of TW solutions to the equation
(\ref{GBE1}), obtained in \cite{Vlask_06}. The structure of the
study is following. In section 2 we present a family of exact TW
solutions, satisfying under certain conditions equation (\ref{GBE1})
and formulate the conditions that guarantee the stability of some
exact solutions in explicit form. In section 3 we construct the
numerical scheme based on the Godunov method
\cite{Godunov}--\cite{Vsan_08} and bring the results of numerical
simulations, backing the qualitative study and partly completing it.
In the last section we briefly summarize the results obtained and
outline the ways of further investigations.

\section{Stability analysis of the exact solutions to the equation (\ref{GBE1})}

\subsection{Statement of the problem}

Let us reformulate the results obtained in \cite{Vlask_06} for the
equation (\ref{GBE1}), assuming that \[f(u)=\nu\,\left(u-m_1
\right)\,\left(u-m_2 \right)\,\left(u-m_3 \right),\] where
$m_k,\,\,\,k=1,\,2,\,3,$ are constant parameters. We are looking for
the TW solutions
\begin{equation}\label{TW_ans}
u(t,\,x)=U\left(z\right)\,\equiv\,U\left(x-V\,t\right),
\end{equation}
where $V$ is a constant velocity of the wave pack. After the formal
substitution of the   travelling wave ansatz (\ref{TW_ans}) into the
equation (\ref{GBE1}), one obtains a nonlinear second order ODE
\[
\left(\alpha\,V^2-\kappa\right)\,\ddot{U}+\dot{U}\,\left(\mu\,U-V\right)=\nu\,\left(U-m_1
\right)\,\left(U-m_2 \right)\,\left(U-m_3 \right),
\]
which is, generally speaking, non-integrable. In order to obtain the
exact TW solutions, we employ a Hirota-like ansatz
$u(t,\,x)=\frac{\Psi'(\xi)}{\Psi(\xi)}$, which, being substituted to
(\ref{GBE1}), leads to the following third-order ODE:
\begin{eqnarray}\label{bilin}
\Psi^2\,\left[\Delta\,\Psi'''-V\,\Psi''-\nu\,\sum_{i\neq j}
m_i\,m_j\Psi'+\nu\,m_1\,m_2\,m_3\,\Psi
 \right] + \nonumber \\
    +\Psi\,\Psi'\left[\left( \mu-3\,\Delta  \right)\Psi'' + \left(V+\nu\,\sum_{i=1}^3{m_i}
   \right)\Psi'\right]+\nonumber \\
   +\left(2 \Delta-\mu-\nu  \right)\left(\Psi' \right)^3 =0, \label{fullhireq}
\end{eqnarray}
where $\Delta=\alpha\,V^2-\kappa.$ For physical reasons
\cite{Landau}, we assume that $\Delta>0$.

On first sight, the equation (\ref{fullhireq}) is even more
complicated than that obtained by the convenient TW ansatz
(\ref{TW_ans}). But as it easily seen,  (\ref{fullhireq}) reduces to
the linear equation
\begin{equation}\label{GBE1_lin}
\Delta\,\Psi'''-V\,\Psi''-\nu\,\left[
\left(m_1\,m_2+m_1\,m_3+m_2\,m_3 \right)\Psi'-m_1\,m_2\,m_3\,\Psi
\right]=0,
\end{equation}
provided that
\begin{eqnarray}
\mu=3\,\Delta, \label{constr_1} \\
\nu=-\Delta,  \label{constr_2} \\
V=\left(m_1+m_2+m_3 \right) \Delta.  \label{constr_3}
\end{eqnarray}
As it was shown in \cite{Vlask_06},  the roots of the characteristic
equation, corresponding to the linear equation (\ref{GBE1_lin}),
coincide with the parameters $m_1,\,\,m_2,\,\,m_3$ when the
restrictions (\ref{constr_1})-(\ref{constr_3}) take place. This
enables us to formulate the following result:

{\bf Theorem 1.} Let the equalities
(\ref{constr_1})-(\ref{constr_3}) take place. Then, depending on the
values of the parameters $\left\{m_k\right\}_{k=1}^3$, the equation
(\ref{GBE1}) has the following exact solutions:

\begin{enumerate}

\item
\begin{equation}\label{m_123}
u(t,\,x)=\frac{m_1\,C_1\,\exp{[m_1\,z]}+m_2\,C_2\,\exp{[m_2\,z]}+m_3\,C_3\,\exp{[m_3\,z]}}
{C_1\,\exp{[m_1\,z]}+C_2\,\exp{[m_2\,z]}+C_3\,\exp{[m_3\,z]}}
\end{equation}
if $m_1\neq\,m_2\,\neq\,m_3\,\neq\,m_1$;

\vspace{2mm}

\item
\begin{equation}\label{m_12}
u(t,\,x)=\frac{m_1\,C_1\,\exp{[m_1\,z]}+\exp{[m_2\,z]}\left[C_2\,m_2+C_3+m_2\,C_3\,z
\right]} {C_1\,\exp{[m_1\,z]}+\exp{[m_2\,z]}\left[C_2+z\,C_3
\right]}
\end{equation}
if $m_1\neq\,m_2\,=\,m_3$;

\vspace{2mm}

\item
\begin{equation}\label{m}
u(t,\,x)=m+\frac{C_2+2\,z} {C_3+z\left(C_2+z  \right)}
\end{equation}
if $m_1=\,m_2\,=\,m_3=m$;

\vspace{2mm}

\item
\begin{equation}\label{m_12_cplx}
u(t,\,x)=\frac{m_3\,C_3\,\exp{[m_3\,z]}+2\,\exp{[\alpha\,\xi]}\,
\left[\alpha\,\cos{(\beta\,z)}-\beta\,\sin{\beta\,z}
 \right]}
 {C_3\,\exp{[m_3\,z]}+2\,\exp{[\alpha\,z]}\,\cos{(\beta\,z)}},
\end{equation}
if $m_3$ is real, while $\bar{m_2}=m_1=\alpha+i\,\,\beta$,
$\alpha,\,\beta\,\,\in\,R$.

\end{enumerate}

\vspace{3mm}

{\bf Remark.} {\it Using (\ref{constr_1})--(\ref{constr_3}), and the
inequality $\Delta>0$, one easily gets the following expression for
the wave pack velocity:
\begin{equation}\label{V_exact}
V=\Delta\,\sum_{i=1}^3{m_i}=\frac{1+\sqrt{1+4\,\alpha\,\kappa\,\left(\sum_{i=1}^3{m_i}
 \right)^2}}{2\,\alpha\,\sum_{i=1}^3{m_i}}.
 \end{equation}
}

 \vspace{3mm}

In order to study the stability of TV solutions, depending in fact
on a single variable $z=t-\,V\,x$, it is instructive to pass to new
independent variables
\[
\bar{t}=t, \qquad \bar{z}=x-V\,t,
\]
in which the invariant solutions (\ref{m_123})--(\ref{m_12_cplx})
become stationary. In the new variables the equation (\ref{GBE1})
reads as follows:
\begin{equation}\label{GBE1B}
\alpha\,\left[\frac{\partial}{\partial\,\bar{t}}-V\frac{\partial}{\partial\,\bar{z}}\right]^2\,u+
\left[\frac{\partial}{\partial\,\bar{t}}-V\frac{\partial}{\partial\,\bar{z}}\right]\,u+
\mu\,u\,\frac{\partial\,u}{\partial\,\bar{z}}-\kappa\,\frac{\partial^2\,u}{\partial\,\bar{z}^2}=f(u)
\end{equation}
 (for simplicity, we omit the bars over the independent variables henceforth).

On studying the stability of stationary solutions, we proceed in the
standard way, presenting the perturbed solution in the form
\begin{equation}\label{perturbed}
u(t,\,x)=U(z)+\epsilon\,\exp{[-\lambda\,t]}\,g(z),
\end{equation}
where $U(z)$ denotes one of the TW solutions described by the
theorem 1. Up to $O(\epsilon^2)$, the function $g(z)$ satisfies the
equation
\begin{eqnarray*}
\Delta\,g''(z)+\left(\mu\,U(z)+2\,\alpha\,V\,\lambda-V \right)
g'(z)+  \\ \left[     \alpha\,\lambda^2 -\nu\,\sum_{i\neq
j}{m_i\,m_j}+2\,\nu\,U(z)\sum_{i=1}^3\,m_i-3\,\nu\,U^2(z)+\mu\,U'(z)\right]\,g(z).
\end{eqnarray*}
For technical reasons, it is instructive to get rid of the terms
proportional to  $g'(z)$, and this can be done by the substitution
\begin{equation}\label{subst_g_h}
g(z)=h(z)\,\exp{[\varphi(z)]}.\end{equation} One easily verifies by
the direct inspection, that the following statement holds true:

\vspace{3mm}

{\bf Lemma 1.} If \[
\varphi'(x)=-\frac{\mu\,U(z)+2\,\alpha\,V\,\lambda-V}{2\,\Delta}, \]
then the function $h(z)$ satisfies the equation
\begin{eqnarray}
\hat{L}\left[h(z),\,\lambda
 \right]= \label{eqfor_h} \\
 =4\,\alpha\,\kappa\,h(z)\,\lambda^2-4\,B(z)\,h(z)\,\lambda+\Delta^2\,h(z)\,K(z)-4\,\Delta^2\,h''(z)=0,
 \nonumber
\end{eqnarray}
where
\begin{equation}\label{exprB}
B(z)=\kappa-3\,\alpha\,\Delta^2\,U(z)\sum_{i=1}^3m_i,
\end{equation}
\begin{equation}\label{exprK}
K(z)= \sum_{i=1}^3m_i^2-2\,\sum_{i\neq
j}\,m_i\,m_j+2\,U(z)\sum_{i=1}^3m_i-3\,U^2(z)-6\,U'(z).
\end{equation}

\vspace{3mm}

So, using the ansatz (\ref{perturbed}), followed by the substitution
(\ref{subst_g_h}), we get the generalized eigenvalue problem
(\ref{eqfor_h}). Evidently, the stability of the self-similar
solution $U(z)$ can be achieved, if all possible values of the
parameter $\lambda$ are positive.

In what follows, we'll restrict our consideration to a family of
perturbations, vanishing beyond some compact set $<-L,\,\,L>$. With
such a restriction, we get the eigenvalue problem
\begin{equation}\label{quasi_Sturm}
\hat{L}\left[h(z),\,\lambda\,\right]=0, \qquad h(-L)=h(L)=0.
\end{equation}
Let us note, that in the parabolic case, i.e., when $\alpha=0$,
(\ref{quasi_Sturm}) reduces to the standard  Sturm-Liouville
Boundary Value Problem. Although the eigenvalue problem we deal with
differs from the classical one, the main conclusions concerning the
properties of the eigenvectors and eigenfunctions remain the same
under quite general assumptions \cite{Keldysh}, satisfied with
certain by the functions $B(z)$ and $K(z)$.

In order to obtain the restrictions on the signs of the eigenvalues
$\lambda$, we multiply the equation (\ref{eqfor_h}) by $h(z)$ and
then integrate the resulting equation over $z$ from $-L$ to $L$. As
a result, we obtain the  quadratic equation with respect to
$\lambda$:
\begin{equation}\label{quadratic}
\lambda^2-b\,\lambda+r=0,
\end{equation}
where
\[
b=\frac{\int_{-L}^L{B(z)h(z)^2\,d\,z}}{\alpha\,\kappa\,||h||^2},
\qquad
r=\frac{\Delta^2\,\int_{-L}^L{K(z)h(z)^2\,d\,z}+4\,\Delta^2\,||h'||^2}{4\,\alpha\,\kappa\,||h||^2},
\]
$||h||^2=\int_{-L}^L{h(z)^2\,d\,z},$
$||h'||^2=\int_{-L}^L{h'(z)^2\,d\,z}.$ From the above formulae, we
get the following relations concerning the roots of the quadratic
equations:
\begin{eqnarray}
\lambda_1+\lambda_2=\frac{\int_{-L}^L{B(z)h(z)^2\,d\,z}}{\alpha\,\kappa\,||h||^2},
\label{lam_12plus}
\\
\lambda_1\,\lambda_2=\Delta^2\,\frac{\int_{-L}^L{K(z)h(z)^2\,d\,z}+4\,||h'||^2}{4\,\alpha\,\kappa\,||h||^2}.
\label{lam_12cdot}
\end{eqnarray}
This immediately leads us to the statement:

{\bf Proposition 1.} In order that the eigenvalues $\lambda_k, \quad
k=1,\,2$ be positive, it is sufficient that the functions $B(z)$ and
$K(z)$, restricted to the segment $<-L,\,L>$, satisfy the following
inequalities:
\begin{equation}\label{sufcond}
B(z)>0, \qquad K(z)\geq 0.
\end{equation}
Below we pose the conditions that guarantee the fulfillment of the
inequalities (\ref{sufcond}) for some  exact invariant solutions to
the equation (\ref{GBE1}).

\subsection{ Stability analysis of the solution (\ref{m_123}) }

We restrict our consideration to the real constants
$\left\{m_i\right\}_{i=1}^3$. Without the loss of generality, we can
assume that  they are ordered as follows:
$0\leq\,m_1\leq\,m_2\leq\,m_3$. We assume in addition that the
constant $C_1$  is nonzero, and the solution (\ref{m_123}) can be
rewritten in the form
\begin{equation}\label{m_different_2}
U(z)=\frac{\Psi'(z)}{\Psi(z)}, \qquad
\Psi(z)=\exp{\left[m_1\,z\right]}+{C_2}\,\exp{\left[m_2\,z\right]}+{C_3}\,\exp{\left[m_3\,z\right]}
\end{equation}
The solution (\ref{m_different_2}) occurs to possess the following
property:

\vspace{3mm}

{\bf Lemma 2. } Function (\ref{m_different_2}) is monotonic for any
positive $C_2$ and $C_3$ and satisfies the inequalities
\begin{equation}\label{estim_1}
m_1 < U(z) < m_3.
\end{equation}
\vspace{3mm}

{\it Proof.} Since the derivative of $U(z)$ is expressed by the
formula \[U'(z)=\frac{\Psi''(z)\Psi(z)-\Psi'(z)^2 }{ \Psi(z)^{2}},\]
we concentrate upon the estimation of the sign of the numerator.
After some algebraic manipulation, performed with the help of {\it
Mathematica} package (and which can be easily verified manually), we
get the inequality
\begin{eqnarray*}
\Psi''(z)\Psi(z)-\Psi'(z)^2=C_2\,\exp{[m_2\,z]}\left\{\exp{[m_1\,z]}\left(m_1-m_2\right)^2+\right.\\\left.+
C_3\,\exp{[m_3\,z]}\left(m_2-m_3\right)^2\right\}
+C_3\,\exp{[(m_1+m_3)\,z]}\left(m_1-m_3\right)^2\,>\,0.
\end{eqnarray*}
The validity of the inequalities (\ref{estim_1}) appear from the
calculation of limits:
\[
\lim\limits_{z\to +\infty}U(z)=m_3, \qquad \lim\limits_{z\to
-\infty}U(z)=m_1.
\]

\vspace{3mm}

The above lemma can be used for the estimation of the signs of
inequalities (\ref{sufcond}). We begin with the first one. The
validity of the inequality $B(z)>0$ appears from the inequality
\[
\kappa>\alpha\,\mu\,V\,\mbox{max}_{z\in
R}U(z)=\alpha\,\mu\,V\,m_3>\alpha\,\mu\,V\,U(z).
\]
Taking into account the conditions
(\ref{constr_1})--(\ref{constr_3}), and expressing $V$ by means of
the formula (\ref{V_exact}) we can rewrite the  inequality
$\kappa>\alpha\,\mu\,V\,m_3$ in the form

\[
\kappa>3\,\alpha\,m_3\frac{\left[1+\sqrt{1+4\,\alpha\,\kappa\,(m_1+m_2+m_3)^2}\right]^2}{4\,\alpha^2\,(m_1+m_2+m_3)^3}
,
\]
or, what is the same,
\begin{eqnarray*}
4\,\alpha\,\kappa\,(m_1+m_2+m_3)^3> \\>3\,m_3\,\left\{
2+4\,\alpha\,\kappa\,(m_1+m_2+m_3)^2+2\,\sqrt{1+4\,\alpha\,\kappa\,(m_1+m_2+m_3)^2}
\right\}. \end{eqnarray*} This, in turn, is equivalent to
\begin{eqnarray}
4\,\alpha\,\kappa\,(m_1+m_2+m_3)^2\left[ (m_1-m_3)+(m_2-m_3)
\right]>\nonumber\\>6\,m_3\,\left\{
1+\sqrt{1+4\,\alpha\,\kappa\,(m_1+m_2+m_3)^2} \right\}.
\label{ineq_M_123}
\end{eqnarray}
It is evident, that under the above assumptions the inequality
(\ref{ineq_M_123}) cannot be fulfilled, so, following this way we
cannot gain any useful information. It occurs to be possible, if we
restrict the set of exact solutions described by the formula
(\ref{m_different_2}) by putting $C_3=0$:
\begin{equation}\label{m_different_3}
U(z)=\frac{m_1\,\exp{[m_1\,z]}+{C_2}\,m_2\,\exp{[m_2\,z]}}{\exp{[m_1\,z]}+{C_2}\,\exp{[m_2\,z]}}.
\end{equation}
In analogy with the lemma 2, one can check the validity of the
following statement:

\vspace{3mm}

{\bf Lemma 3.} if $C_2>0$, then the function (\ref{m_different_3})
is monotonic and satisfies the inequalities
\[
m_1<U(z)<m_2.
\]
So, now there is the  condition $\kappa-\alpha\,\mu\,V\,m_2>0$,
which guarantees the validity of the inequality $B(z)>0$ for the
exact solution (\ref{m_different_3}). Using
(\ref{constr_1})-(\ref{constr_3}), and their consequence
(\ref{V_exact}), we get the inequality
\begin{eqnarray}\label{m_different_ineq}
4\,\alpha\,\kappa\,(m_1+m_2+m_3)^2\left[ (m_1-m_2)+(m_3-m_2)
\right]>\nonumber\\>6\,m_2\,\left\{
1+\sqrt{1+4\,\alpha\,\kappa\,(m_1+m_2+m_3)^2} \right\}.
\end{eqnarray}
Obviously, the inequality (\ref{m_different_ineq}) does not hold for
arbitrary values of the parameters. Yet if all the parameters, but
$m_3$ are fixed, then the LHS behaves as $m_3^3$ while the RHS as
$m_3^1.$ Hence there exists the critical value $m_3^{*_1}$ such that
for any $m_3\geq m_3^{*_1}$ the inequality (\ref{m_different_ineq})
does take place.

Now let us consider the condition
\begin{eqnarray*}
K(z)= \sum_{i=1}^3m_i^2-2\,\sum_{i\neq
j}\,m_i\,m_j+2\,U(z)\sum_{i=1}^3m_i-3\,U^2(z)-6\,U'(z)\geq 0.
\end{eqnarray*}
Below we shall use the following elementary statement:

\vspace{3mm}
  {\bf Lemma 4.}
The derivative of the function (\ref{m_different_3}) satisfies the
inequality
\[
0<U'(z)\leq \frac{(m_1-m_2)^2}{4}.
\]
Using the above restrictions and the statements of the lemma 3, we
get the estimation
\begin{eqnarray}
K(z)>\sum_{k=1}^{3}m_k^2-2\sum_{i\neq j}m_i\,m_j+ \label{m_different_ineq2}\\
+2\,m_1\,\sum_{k=1}^{3}m_k-3\,m_2^2-6\frac{(m_2-m_1)^2}{4}=K_1.
\nonumber
\end{eqnarray}
Fixing all the parameters but $m_3$, we can treat $K_1$ as the
quadratic function:
\[K_1=m_3^2-2\,m_2\,m_3^1+C\left(m_1,\,m_2\right).\]  So there exists
a number $m_3^{*_2}$ such that for $m_3 > m_3^{*_2},$ $K_1$ is
positive. From this immediately follows the main result of this
section:

\vspace{3mm}

{\bf Theorem 2.} If
$m_3>\,\,\mbox{max}\left\{m_3^{*_1},\,\,m_3^{*_2}\right\}$, then,
under the restrictions stated above, the TW solution
(\ref{m_different_3}) is stable.

\vspace{3mm}

\section{Numerical study of the invariant TW solutions to the equation (\ref{GBE1})}

\subsection{Construction  of the numerical scheme.}
We base our numerical calculations on the Godunov method
\cite{Godunov,RozhdYan}. It is not difficult to extend the
construction of a numerical scheme upon somewhat more general
equation
\begin{equation}\label{GBE2}
\alpha\,u_{tt}+u_t+\mu\,u\,u_x-\kappa\,
\left[u^n\,u_{x}\right]_{x}=f(u),
\end{equation}
containing  nonlinear diffusive term. Introducing the new variable
$\Psi=u_t-\sqrt{\kappa\,\gamma\,u^n}\,u_x,\,\,\,\gamma=\alpha^{-1}$,
we can rewrite the equation (\ref{GBE2}) in the form of the first
order system:
\begin{equation}\label{GBE_sys_1}
\frac{\partial\,}{\partial\,t}\left(\begin{array}{c}u \\ \Psi
\end{array}\right)+\left(\begin{array}{cc}-\sqrt{\kappa\,\gamma\,u^n} & 0 \\
\gamma\,u+\Psi\,\sqrt{\gamma}\,\frac{n}{2}u^{n/2-1}+\kappa^{1/2}\gamma^{3/2}u^{n/2}
& \sqrt{\kappa\,\gamma\,u^n}
\end{array}\right)\frac{\partial\,}{\partial\,t}\left(\begin{array}{c}u \\ \Psi
\end{array}\right)=H,
\end{equation}
where $H=\left\{ \Psi,\,\,\gamma\left[f(u)-\Psi\right] \right\}^{\rm
tr},$  and $(\cdot)^{{\rm tr}}$ stands for the operation of
transposition.

Let us consider the calculating cell $a\,b\,c\,d$ (see Fig.~1) lying
between $m-th$ and $(m+1)-th$ temporal layers of the uniform
rectangular mesh.  It is easy to see that the system
(\ref{GBE_sys_1}) can be presented in the following vector form:
\begin{equation}\label{GBE_cons}
\frac{\partial\,F}{\partial\,t}\,+\frac{\partial\,G}{\partial\,x}=H,
\end{equation}
with  $F=\left( u,\,\Psi  \right)^{\rm tr},$
\[
G=\left( -\sqrt{\gamma\,\kappa}\frac{u^{n/2+1}}{n/2+1};\,\,
\frac{\gamma}{2}\left[u^2+\frac{2\,\sqrt{\gamma\,\kappa}}{n/2+1}u^{n/2+1}
\right]+ \Psi\,\sqrt{\gamma\,\kappa\,u^2} \right)^{\rm tr}.
\]
 From (\ref{GBE_cons}) arises the
equality of integrals
\[
\int\int_{\Omega}\,{\left(\frac{\partial\,F}{\partial\,t}\,+\frac{\partial\,G}{\partial\,x}\right)}dx\,dt=
\int\int_{\Omega}\,{H}dx\,dt,
\]
where $\Omega$ is identified with the rectangle  $a\,b\,c\,d$. Due
to the Gauss-Ostrogradsky theorem, integral in the LHS can be
presented in the form
\begin{equation}\label{Gauss}
\int\int_{\Omega}\,{\left(\frac{\partial\,F}{\partial\,t}\,+\frac{\partial\,G}{\partial\,x}\right)}dx\,dt=
\oint_{\partial\,\Omega}{G\,d\,t-F\,d\,x}.
\end{equation}

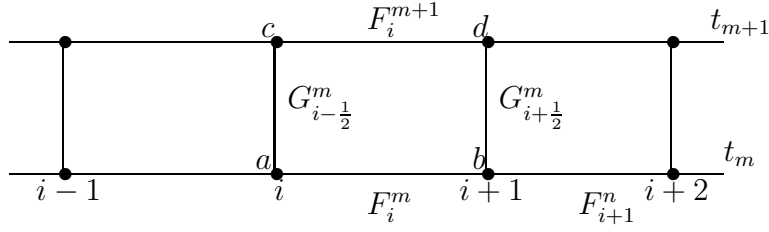
\begin{figure}
\begin{picture}(380,80)
\put(70,50){\line(4,0){270}} \put(70,0){\line(4,0){270}}
 \put(90,0){\line(0,4){50}} \put(170,0){\line(0,4){50}}
\put(250,0){\line(0,4){50}} \put(320,0){\line(0,4){50}}
\put(163,2){$a$} \put(245,2){$b$} \put(165,52){$c$}
\put(245,52){$d$} \put(170,-10){$i$} \put(240,-10){$i+1$}
\put(310,-10){$i+2$} \put(80,-10){$i-1$}\put(88,-3){$\bullet$}
\put(88,47){$\bullet$} \put(168,-3){$\bullet$}
\put(168,47){$\bullet$} \put(248,-3){$\bullet$}
\put(248,47){$\bullet$} \put(318,-3){$\bullet$}
\put(318,47){$\bullet$}
 \put(205,-15){$F_{i}^m$}
\put(285,-15){$F_{i+1}^n$} \put(205,55){$F_{i}^{m+1}$}
\put(175,25){$G_{i-\frac{1}{2}}^m$}
\put(255,25){$G_{i+\frac{1}{2}}^m$} \put(340,5){$t_m$}
\put(335,55){$t_{m+1}$}

\end{picture}
\vspace{10pt} \caption{Scheme of the calculating cell}\label{fig:1}
\end{figure}

\noindent
 Let us denote the distance between  the $i-th$ and
$(i+1)-th$ nodes of the $OX$ axis by $\Delta\,x$ while the distance
between the two adjacent temporal layers by $\Delta\,t$. Then, up to
$O\left(\left|\Delta\,x\right|^2,\,\,\left|\Delta\,t\right|^2\right)$,
we get from the equation (\ref{GBE_cons})   the following difference
scheme:
\begin{equation}\label{Num_vec}
\left(F_i^{m+1}- F_i^{m}
\right)\,\Delta\,x+\left(G_{i+\frac{1}{2}}^{m}-
G_{i-\frac{1}{2}}^{m}
\right)\,\Delta\,t=H_i^{m}\Delta\,t\,\Delta\,x,
\end{equation}
where $G_{i+\frac{1}{2}}^{m},\,\,G_{i-\frac{1}{2}}^{m}$  are the
values of the vector-function $G$  on the segments $b\,d$ and
$a\,c$, correspondingly. In the Godunov method these values are
defined by solving the Riemann problem. Below we describe the
procedure of their calculation.

\begin{figure}
\begin{picture}(380,160)
\put(200,0){\line(1,1){110}} \put(200,0){\line(-1,1){110}}
\put(200,0){\vector(0,4){120}} \put(70,0){\line(4,0){270}}
\put(150,0){\vector(1,1){40}} \put(230,0){\vector(-1,1){40}}
\put(150,20){$r_{+}$} \put(210,25){$r_{-}$} \put(150,20){$r_{+}$}
\put(110,40){$I$} \put(210,25){$r_{-}$} \put(270,40){$III$}
\put(180,50){$II$} \put(210,25){$r_{-}$} \put(190,120){$t$}
\put(210,120){$x=0$} \put(350,0){$x$} \put(80,15){$u_1,\,\,\Psi_1$}
\put(265,15){$u_2,\,\,\Psi_2$} \put(300,120){$x=C_1\,t$}
\put(130,95){$u_{II},\,\,\Psi_{II}$} \put(60,120){$x=-C_1\,t$}
\end{picture}
\vspace{5pt} \caption{Scheme of solving  the Riemann
problem}\label{fig:2}
\end{figure}
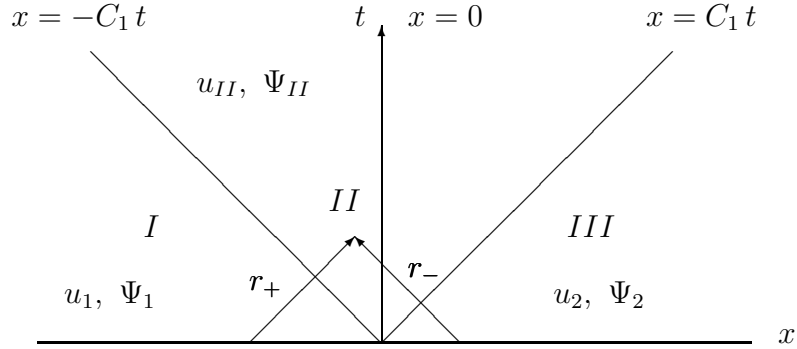

In accordance with common practice, instead of dealing with the
initial system (\ref{GBE_sys_1}), we look for the solution of the
Riemann problem $( u_1,\,\, \Psi_1)$ at $x<0$ and $(u_2,\,\,
\Psi_2)$ at $x>0$ to corresponding homogeneous system
\begin{eqnarray}\label{homogsys}
\frac{\partial\,}{\partial\,t}\left(\begin{array}{c}u \\ \Psi
\end{array}\right)+\left(\begin{array}{cc}-C_1 & 0 \\
C_2 & C_1
\end{array}\right)\frac{\partial\,}{\partial\,t}\left(\begin{array}{c}u \\ \Psi
\end{array}\right)=0,
\end{eqnarray}
where
\begin{eqnarray*}
C_1=\sqrt{\gamma\,\kappa\,u_0^n}, \\
C_2=\gamma\,u_0+\Psi_0\,\sqrt{\gamma}\,\frac{n}{2}u_0^{n/2-1}+\kappa^{1/2}\gamma^{3/2}u_0^{n/2}.
\end{eqnarray*}

Using the linearized system (\ref{homogsys}), it is easy to
calculate the Riemann invariants \[ r_{+}=C_2\,u+2\,C_1\,\Psi,
\qquad r_{-}=u,
\]
corresponding to the characteristic velocities $C_{\pm}=\pm\,C_1$.
Characteristics $x=\pm C_1\,t$ divide the half-plane $t\geq 0$ into
three sectors (see Fig.~\ref{fig:2}) and the problem is to find the
values of the parameters in sector II, basing at the values
$(u_1,\,\Psi_1)$  and $(u_2,\,\Psi_2)$, which are assumed to be
defined. The scheme of calculating the values $u_{II},\,\,\Psi_{II}$
is based on the property of the Riemann invariants to retain their
values along the corresponding characteristics. From this we get the
system of algebraic equations
\begin{eqnarray*}
C_2\,u_1+2\,C_1\,\Psi_1=C_2\,u_{II}+2\,C_1\,\Psi_{II},  \\
u_2=u_{II}. \nonumber
\end{eqnarray*}
So the values of the parameters $u,\,\Psi$ in the sector
$-C_1\,t<x<C_1\,t$ are given by the formulae:
\begin{eqnarray}\label{acous_appr}
  u_{II}=u_2,
\\  \Psi_{II}=\Psi_{1}+C_2\,\frac{u_1-u_2}{2\,C_1}. \nonumber
\end{eqnarray}

Thus, the difference scheme for (\ref{GBE2}) takes the following
form:
\begin{eqnarray*}
 u_i^{m+1}=u_i^m+\frac{\Delta\,t}{\Delta\,x}
 \left[
 \left(G_1\right)_{i-\frac{i}{2}}^m-\left(G_1\right)_{i+\frac{i}{2}}^m
 \right]+{\Delta\,t}\,\left( H_1\right)_i^m, \\
 \Psi_i^{m+1}=\Psi_i^{m}+\frac{\Delta\,t}{\Delta\,x}
 \left[
 \left(G_2\right)_{i-\frac{i}{2}}^m-\left(G_2\right)_{i+\frac{i}{2}}^m
 \right]+{\Delta\,t}\,\left( H_2\right)_i^m,
\end{eqnarray*}
where
\begin{eqnarray*}
\left(G_1\right)_{i-\frac{1}{2}}^m=
-\sqrt{\gamma\,\kappa}\frac{\left(u_{i-\frac{1}{2}}^m\right)^{n/2+1}}{n/2+1},
\quad i=2,3,...,N-1,
\\
\left(G_2\right)_{i-\frac{1}{2}}^m=\frac{\gamma}{2}\left[\mu\left(u_{i-\frac{1}{2}}^m\right)^2+
\frac{2\,\sqrt{\gamma\,\kappa}}{n/2+1}\left(u_{i-\frac{1}{2}}^m\right)^{n/2+1}
\right]+
\left(\Psi_{i-\frac{1}{2}}^m\right)\,\sqrt{\gamma\,\kappa\,\left(u_{i-\frac{1}{2}}^m\right)^{n/2}}
,
\end{eqnarray*}
$\left(u_{i-\frac{1}{2}}^m\right),\,\,\left(\Psi_{i-\frac{i}{2}}^m\right),\,\,i=2,3,...N-1,
$ are calculated by means of the formula (\ref{acous_appr}), in
which $(u_1,\,\Psi_1)$ and $(u_2,\,\Psi_2)$ are substituted,
correspondingly, by $(u_{i-1}^m,\,\Psi_{i-1}^m)$ and
$(u_{i}^m,\,\Psi_{i}^m)$,  while the constants $C_k,\,\,k=1,2$ take
the form
\begin{eqnarray*}
C_1=\sqrt{\gamma\,\varkappa\,\left(u_{i-1}^m\right)^n}, \\
C_2=\gamma\,\left(u_{i-1}^m\right)+\left(\Psi_{i-1}^m\right)\,\sqrt{\gamma}\,
\frac{n}{2}\left(u_{i-1}^m\right)^{n/2-1}+\varkappa^{1/2}\gamma^{3/2}\left(u_{i-1}^m\right)^{n/2}.
\end{eqnarray*}

\subsection{Results of numerical simulation}

\begin{figure}
 \centering\includegraphics[width=3 in, height=2 in ]{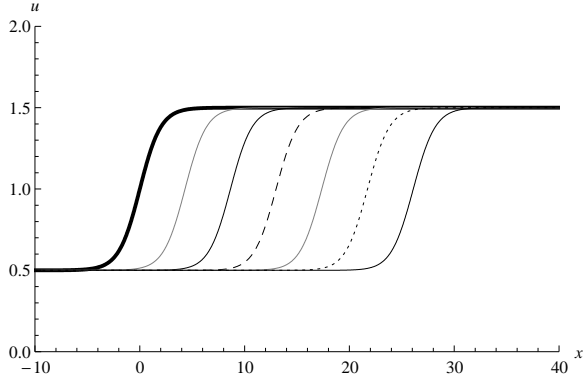}
\caption{Numerical solution  of the system (\ref{GBE1}) in case when
the invariant  kink-like solution (\ref{m_different_3}) with
$\alpha=\kappa=1,\,\, m_1=0.5,\, m_2= 1.5,\, m_3 = 5,\, C_1=C_2=1,\,
C_3=0 $ is taken as the Cauch\'{y} data. Successive graphs of TW
solution, moving from left to right, correspond to
$t_i=4\,(i-1),\,\,\,i=1,...7$}\label{fig:U123C30farA}
\end{figure}

\begin{figure}
 \centering\includegraphics[width=3 in, height=2 in ]{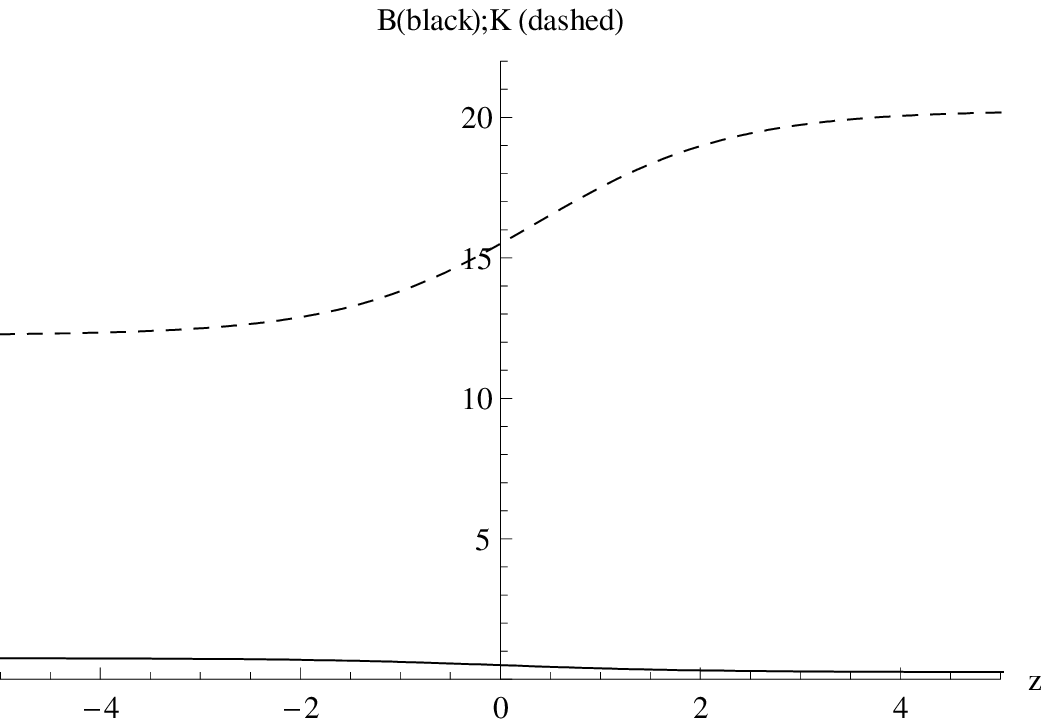}
\caption{Graphs of the functions $B(z)$ and $K(z)$, obtained for
$\alpha=\kappa=1,\, m_1=0.5,\, m_2= 1.5,\, m_3 = 5,\,
C_1=C_2=1,\,\,\, \mbox{and}\,\, C_3=0 $ }\label{fig:BKstab}
\end{figure}

Below we present the results of numerical solution of the Cauch\'{y}
problem for system (\ref{GBE1}).  In all numerical experiments the
parameters $\alpha$ and $\kappa$ were taken to be equal to one,
while the remaining parameters varied from one case to another.  In
the first series of the numerical experiments we  put $m_1=0.5$,
$m_2=1.5$, and, since for this choice  $m_3^*\approx 4.4$, we took
 $m_3=5$ in order to satisfy the requirements of the theorem 2. As the Cauch\'{y} data we
used the invariant solution described by the formula
(\ref{m_different_3}), and corresponding to $t_0=0.$ Results of the
numerical simulation are shown in Fig.~\ref{fig:U123C30farA}. It is
seen that the kink-like solution evolves for a long time in a stable
self-similar mode. Figure \ref{fig:BKstab} shows the graphs of the
functions $B(z)$ and $K(z)$ for the above values of the parameters.
It can be seen on this graphs that both of the functions are
strictly positive in a vicinity of the front of the kink-like
solution (\ref{m_different_3}).

\begin{figure}
 \centering\includegraphics[width=3 in, height=2 in ]{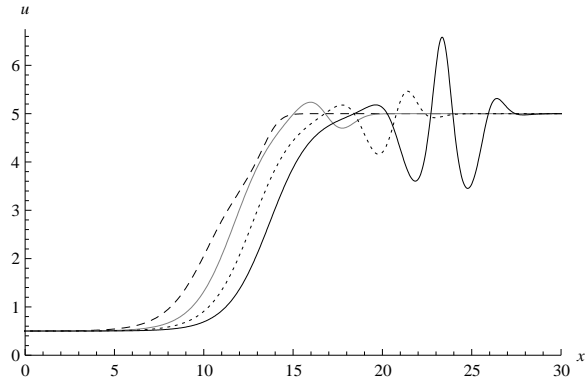}
\caption{Numerical solution  of the system (\ref{GBE1}) in case when
the invariant  kink-like solution (\ref{m_123}) with
$\alpha=\kappa=1,\, m_1=0.5,\, m_2= 1.5,\, m_3 = 5,\, C_1=C_2=1,\,
C_3=3 $ is taken as the Cauch\'{y} data. Successive graphs
correspond to $t_i=12+4\,(i-1),\,\,\,i=1,...4$}\label{fig:U123farB}
\end{figure}

\begin{figure}
 \centering\includegraphics[width=3 in, height=2 in ]{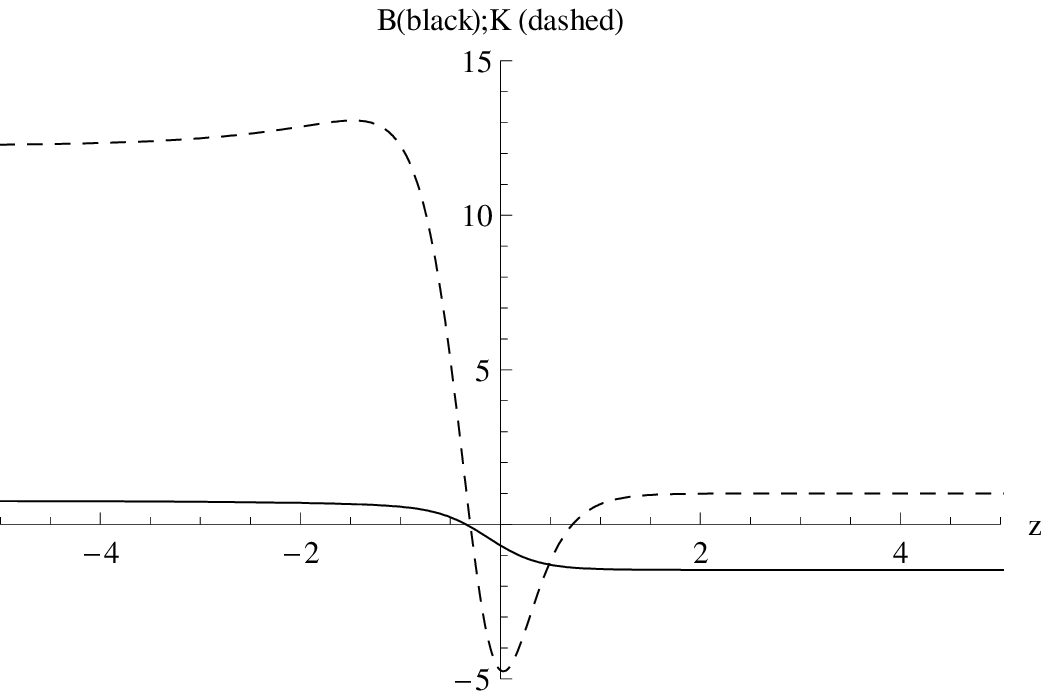}
\caption{Graphs of the functions $B(z)$ and $K(z)$, obtained for
$\alpha=\kappa=1, \,m_1=0.5,\, m_2= 1.5,\, m_3 = 5,\,
C_1=C_2=1,\,\,\, \mbox{and}\,\, C_3=3 $}\label{fig:BK1unst}
\end{figure}

Next we performed the numerical experiments in which the full
solution (\ref{m_123}) was taken as the Cauchy data. We put in this
case $C_3=3$ and the rest of the parameters remained the same as in
the previous case. The results of numerical simulation show that the
self-similar solution is unstable,  Fig.~\ref{fig:U123farB}. The
source of the instability is seen in Fig.~\ref{fig:BK1unst}, showing
the graphs of the functions $B(z)$ and $K(z)$. Both of these
functions are negative in some vicinity of the origin. Besides, the
function $K(z)$ has a local minimum in this vicinity, in  which it
attains a sufficiently large negative value. Analysis of the formula
(\ref{exprK}) shows that the presence of such maximum can be
attributed to the abruptness of the slope of the kink-like solution.
Note that the graphs of the functions shown on
Fig.~\ref{fig:U123farB} correspond to sufficiently large times $t_i
\geq 12,$ for which the effects of instability become evident.
Therefore they do not coincide with initial profile described by the
formula (\ref{m_123}), which is quite sharp.

\begin{figure}
 \centering\includegraphics[width=3.5 in, height=2.25 in ]{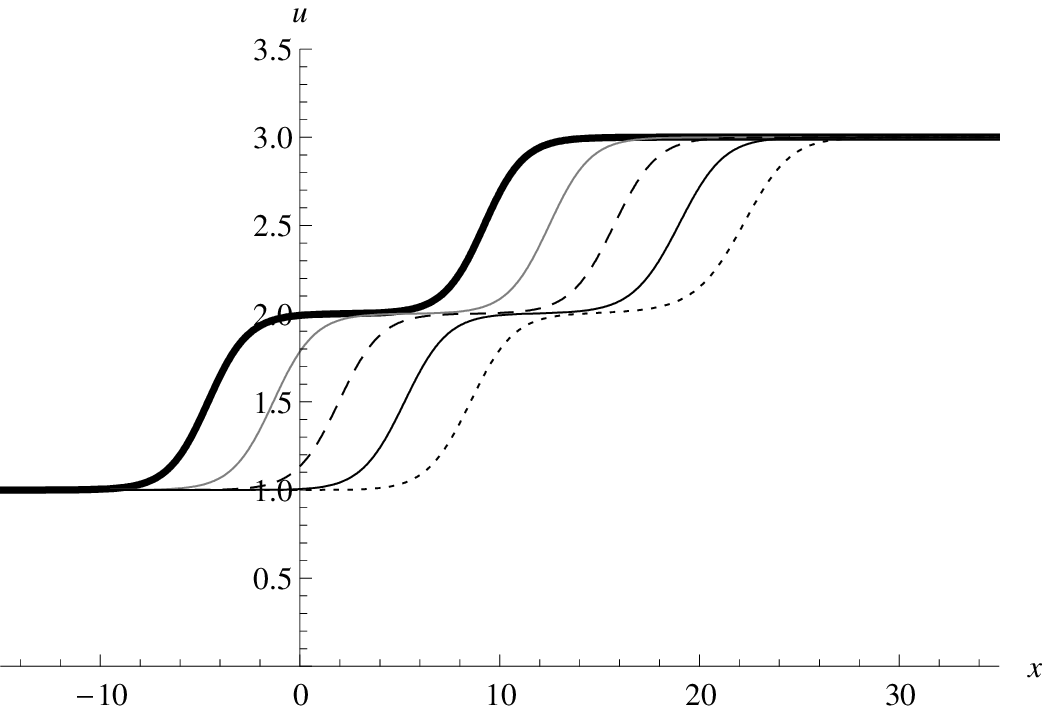}
\caption{Numerical solution of the system (\ref{GBE1}) in case when
the invariant  kink-like solution (\ref{m_123}) with
$\alpha=\kappa=1,\, m_1=1,\, m_2= 2,\, m_3 = 3,\,
C_1=1,\,\,C_2=100,\, C_3=0.01 $ is taken as the Cauch\'{y} data.
Successive graphs of TW solution, moving from left to right,
correspond to $t_i=3\,(i-1),\,\,\,i=1,...5$}\label{fig:U123stepB}
\end{figure}

\begin{figure}
 \centering\includegraphics[width=3 in, height=2 in ]{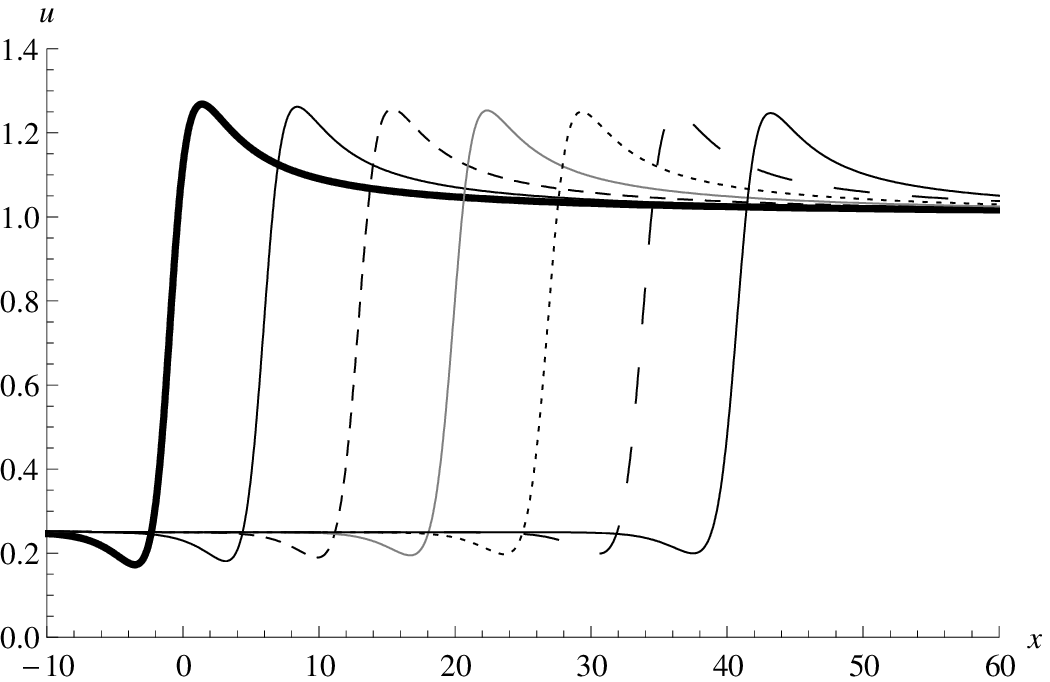}
\caption{Numerical solution of the system (\ref{GBE1}) in case when
the invariant  kink-like solution (\ref{m_12}) with
$\alpha=\kappa=1,\, m_1=0.25,\, m_2=m_3 = 1,\, C_1=C_2=C_3=1 $ is
taken as the Cauch\'{y} data. Successive graphs of TW solution,
moving from left to right, correspond to
$t_i=5.5\,(i-1),\,\,\,i=1,...7$}\label{fig:m122B}
\end{figure}

\begin{figure}
 \centering\includegraphics[width=3 in, height=2 in ]{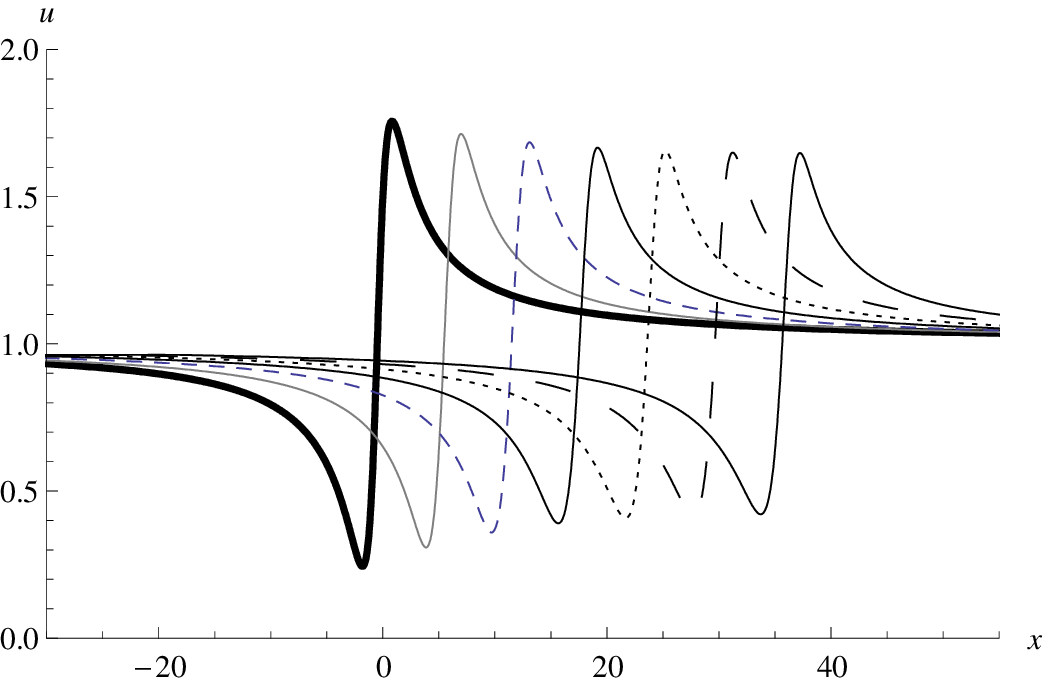}
\caption{Numerical solution of the system (\ref{GBE1}) in case when
the invariant  kink-like solution (\ref{m}) with $\alpha=\kappa=1,\,
m=1,\, C_2=1,\, C_3=2 $ is taken as the Cauch\'{y} data. Successive
graphs of TW solution, moving from left to right, correspond to
$t_i=3.75\,(i-1),\,\,\,i=1,...7$}\label{fig:m111B}
\end{figure}

Let us briefly describe the results of some other numerical
experiments.  Fig.~\ref{fig:U123stepB} shows the results of the
numerical evolution of a step-like initial perturbation described by
the formula (\ref{m_123}).  Fig.~\ref{fig:m122B} shows the results
of the numerical evolution of the initial perturbation described by
the formula (\ref{m_12}). Finally, the Fig.~(\ref{fig:m111B})
describes the numerical evolution of the N-shaped soliton with the
heavy "tail", described by the formula (\ref{m}). Results of
numerical simulation show that the first two wave patterns evolve
without any drastic changes of their shapes. In the last case
diminishing of the maximal and minimal amplitudes is evidently seen.
Besides the wave instability, this effect can be caused by the
numerical scheme's viscosity.

\section{Summary}

So in this paper stability of  TW solutions, satisfying  the
equation  (\ref{GBE1}) under some restrictions on the values of the
parameters are analyzed, and sufficient conditions for the stability
of some family of exact solutions are presented in explicit form. It
is quite evident, that the stability conditions stated by the
theorem 2 are sufficient, but not necessary. In fact, they are the
most strong among all possible conditions of this sort. No wonder,
thus, that for some invariant TW solutions which do not satisfy the
inequalities (\ref{sufcond}), we succeeded to observe the stable
self-similar evolution, as well. To gain the theoretical
justification of the  stability of the exact solutions differen from
(\ref{m_different_3}), an extra qualitative investigations, based on
the more subtle methods are needed.

\vspace{2mm}

{\bf Acknowledgements.}

\vspace{2mm}

\noindent The authors express their gratitude to Prof. Petru
Cojuchari for many fruitful discussions.

\vspace{2mm} \noindent
 This research was supported by the AGH local
grant.

\end{document}